\documentclass{ifacconf}

\usepackage{graphicx}      
\usepackage{natbib}        
\usepackage{amsmath}
\usepackage{amssymb}
\usepackage{mathrsfs}
\usepackage{mathtools}
\usepackage{bm}
\usepackage{cite}
\usepackage{subfigure}

\newcommand{\figref}[1]{Fig.~\ref{#1}}

\newcommand{\thmref}[1]{Theorem~\ref{#1}}
\newcommand{\defref}[1]{Definition~\ref{#1}}
\newcommand{\secref}[1]{Section~\ref{#1}}

\newcommand{\R}{\mathbb{R}}
\newcommand{\Z}{\mathbb{Z}}

\newcommand{\KeyGen}{\mathsf{KeyGen}}
\newcommand{\Enc}{\mathsf{Enc}}
\newcommand{\Dec}{\mathsf{Dec}}
\newcommand{\Eval}{\mathsf{Eval}}
\newcommand{\EC}{\mathsf{EC}}
\newcommand{\pk}{\mathsf{pk}}
\newcommand{\sk}{\mathsf{sk}}
\newcommand{\ct}{\mathsf{ct}}

\newcommand{\tr}{\mathop{\mathrm{tr}}\limits}
\renewcommand{\Vec}{\mathop{\mathrm{vec}}\limits}

\newcommand{\argmin}{\mathop{\mathrm{arg~min}}\limits}

\newcommand{\EV}{\mathop{\mathbb{E}}\limits}

\allowdisplaybreaks[4]
\begin{document}

\thispagestyle{empty}
\fbox{
\begin{minipage}{\textwidth-7mm}\scriptsize
        This work has been submitted to IFAC for possible publication.
    \end{minipage}
}
\newpage
\setcounter{page}{0}
\begin{frontmatter}

\title{Sample Identifying Complexity of Encrypted Control Systems Under Least Squares Identification\thanksref{footnoteinfo}} 

\thanks[footnoteinfo]{This work was supported by JSPS Grant-in-Aid for JSPS Fellows Grant Number JP21J22442 and JSPS KAKENHI Grant Number JP22H01509.}

\author[First,Second]{Kaoru Teranishi}
\author[First]{Kiminao Kogiso}

\address[First]{Department of Mechanical and Intelligent Systems Engineering, The University of Electro-Communications, 1-5-1 Chofugaoka, Chofu, Tokyo 182-8585, Japan (e-mail: teranishi@uec.ac.jp, kogiso@uec.ac.jp)}
\address[Second]{Research Fellow of Japan Society for the Promotion of Science}

\begin{abstract}                
A sample identifying complexity has been introduced in the previous study to capture an adversary's estimation error of system identification.
The complexity plays a crucial role in defining the security of encrypted control systems and designing a controller and security parameter for the systems.
This study proposes a novel sample identifying complexity of encrypted control systems under an adversary who identifies system parameters using a least squares method.
The proposed complexity is characterized by a controllability Gramian and ratio of identification input variance to the noise variance.
We examine the tightness of the proposed complexity and its changes associated with the Gramian and variance ratio through numerical simulations.
The simulation results demonstrate that the proposed complexity captures a behavior of estimation error with a sufficient level.
Moreover, it confirmed that the effect of controllability Gramian in the proposed complexity becomes larger as the variance ratio increases.
\end{abstract}


\end{frontmatter}


\section{Introduction}
\label{sec:introduction}

Outsourcing computation of control systems to a cloud server, such as control as a service (CaaS), is one form of realization of cyber-physical systems that improve the efficiency and flexibility of traditional control systems.
However, such computing services often face threats that adversaries eavesdrop and learn about private information of control systems.
Homomorphic encryption is the major countermeasure against such threats because it provides direct computation on encrypted data without accessing the original messages~\citep{acar2019}.
The encryption was applied to realize an encrypted control that is a framework for secure outsourcing computation of control algorithms~\citep{kogiso2015,farokhi2017,kim2016,kim2022,darup2021}.
Owning to the benefits of encrypted control, various controls, such as model predictive control~\citep{alexandru2018,darup2018a}, motion control~\citep{qiu2019,shono2022}, and reinforcement learning~\citep{suh2021} were implemented in encrypted forms.

Although many encrypted control methods were proposed, it is not sufficiently clarified how secure an encrypted control system is against what type of adversary.
In order to solve this problem, some recent studies have tried to define and analyze the security of encrypted control systems through two approaches.
One of them is a cryptographic approach that defines the provable security of encrypted controls and reveals a relation between the security and existing security notions in cryptography~\citep{teranishi2022a}.
In this security definition, an adversary and information used for attacks are respectively formulated as a probabilistic polynomial-time algorithm and its inputs instead of assuming specific attacks.
Using the security notion, we can analyze qualitative security for a broad class of encrypted control systems.
In contrast, another study has employed a control theoretic approach that considers the security of encrypted control systems under an adversary who wants to learn the system parameters by system identification~\citep{teranishi2022}.
The security in this approach is defined by the system identification error and computation time for the process.
Unlike the cryptographic approach, the security notion in this approach enables quantifying a security level of encrypted control systems.
The study has also solved an optimization problem for designing a controller and security parameter to minimize the computation costs of encryption algorithms while satisfying the desired security level.
Meanwhile, this approach depends on a specific attack scenario: which parameters are identified and what algorithm is used for the identification.

This study proposes a novel sample identifying complexity of encrypted control systems under an adversary who attempts system identification of a plant in order to extend the application of the control-theoretic security analysis.
Although the previous work has focused on an adversary identifying a system matrix of a closed-loop system~\citep{teranishi2022}, this study considers an adversary estimating the system and input matrices of a plant controlled by an encrypted controller.
Such an adversary represents an eavesdropper executing man-in-the-middle attacks and a malicious server infected by malware or spoofing an authorized server computing encrypted control algorithms.
Furthermore, the adversary employs a basic least squares identification method, which is more prevalent in practical use than the Bayesian estimation method discussed in the previous study~\citep{teranishi2022}.
The proposed sample identifying complexity is characterized by a controllability Gramian of the identified plant and variance ratio of adversarial input for the system identification and plant noise.

The proposed sample identifying complexity is beneficial for evaluating the adversary's capability and essential for defining the security of encrypted control systems.
Using the proposed quantity, we can estimate how precisely the adversary is expected to identify a given plant for a certain number of data.
Moreover, the proposed sample identifying complexity is extended to a closed-loop case.
The extended result suggests a defense policy that a controller should be designed to maximize the stability degree of a closed-loop system to prevent system identification.
Our analysis also reveals that such a defense policy is effective when the variance of plant noise is sufficiently smaller than the variance of adversarial identification input.

The rest of this paper is organized as follows.
\secref{sec:preliminaries} defines the syntax of homomorphic encryption and encrypted control and introduces the security definition of encrypted control systems.
\secref{sec:attack_model} formulates an attack model of this study.
\secref{sec:sic} proposes a sample identifying complexity under the attack model.
\secref{sec:simulation} presents the results of numerical simulations.
\secref{sec:conclusion} describes the conclusions and future work.

\section{Preliminaries}
\label{sec:preliminaries}

\subsection{Notation}
\label{sec:notation}

The sets of real numbers and integers are denoted by $\R$ and $\Z$, respectively.
Define the set $\Z^+\coloneqq\{z\in\Z\mid 0\le z\}$.
The sets of $n$-dimensional vectors and $m$-by-$n$ matrices of which elements and entries belong to a set $\mathcal{A}$ are denoted by $\mathcal{A}^n$ and $\mathcal{A}^{m\times n}$, respectively.
The $i$th element of a vector $v\in\mathcal{A}^n$ is denoted by $v_i$.
The induced $2$-norm and Frobenius norm of $M\in\mathcal{A}^{m\times n}$ are denoted by $\|M\|_2$ and $\|M\|_F$, respectively.
The column stack vector of $M$ is defined as $\Vec(M)\coloneqq[M_1^\top\,\cdots\,M_n^\top]^\top$, where $M_i$ is the $i$th column vector of $M$.

\subsection{Homomorphic encryption and encrypted control}
\label{sec:he_ec}

This section defines the syntax of homomorphic encryption and encrypted control.

\begin{defn}\label{def:he}
    Homomorphic encryption is a tuple $\Pi=(\KeyGen,\Enc,\Dec,\Eval)$, where the algorithms are defined as follows.
    \begin{itemize}
        \item $(\pk,\sk)\gets\KeyGen(1^\lambda)$: A key generation algorithm takes a security parameter $1^\lambda$ and outputs a public key $\pk$ and secret key $\sk$, where $1^\lambda$ is the unary representation of $\lambda>0$.
        \item $\ct\gets\Enc(\pk,m)$: An encryption algorithm takes $\pk$ and a plaintext $m$ and outputs a ciphertext $\ct$.
        \item $m\gets\Dec(\sk,\ct)$: A decryption algorithm takes $\sk$ and a ciphertext $\ct$ and outputs a plaintext $m$.
        \item $\ct\gets\Eval(f,\ct_1,\dots,\ct_N)$: An evaluation algorithm takes a function $f:(m_1,\dots,m_N)\mapsto f(m_1,\dots,m_N)$ and ciphertexts $\ct_1,\dots,\ct_N$ and outputs a ciphertext $\ct=\Enc(\pk,f(m_1,\dots,m_N))$, where $\ct_i=\Enc(\pk,m_i)$ for $i=1,\dots,N$.
    \end{itemize}
\end{defn}

For a vector (matrix) plaintext and ciphertext, the algorithms are assumed to perform each element of the vectors (matrices).

An encrypted controller is defined based on the evaluation algorithm of homomorphic encryption as follows.

\begin{defn}\label{def:ec}
    An encrypted controller $\EC$ of controller $f:(\Phi,\xi)\mapsto\psi$ is an algorithm defined as follows, where $\Phi$ is a controller parameter, $\xi$ is a controller input, and $\psi$ is a controller output.
    \begin{itemize}
        \item $\ct_\psi\gets\EC(f,\ct_\Phi,\ct_\xi)$: An encrypted controller algorithm takes a controller $f$ and ciphertexts $\ct_\Phi=\Enc(\pk,\Phi)$ and $\ct_\xi=\Enc(\pk,\xi)$ and outputs a ciphertext $\ct_\psi=\Enc(\pk,f(\Phi,\xi))$ using the evaluation algorithm $\Eval$ in \defref{def:he}.
    \end{itemize}
\end{defn}

The controller parameter and input need to be encoded to plaintexts before encryption because control systems typically operate over real numbers.
Although the encoding causes quantization errors, we ignore the errors for simplicity.
Note that this is the worst case scenario for a defender.

\subsection{Security of encrypted control systems}

This section introduces the security definition of encrypted control systems proposed in the previous study~\citep{teranishi2022}.
The security is defined for encrypted control systems under an adversary who performs system identification of the system by using two notions, sample identifying complexity and sample deciphering time.

A sample identifying complexity is defined for capturing how the expectation of estimation error of the system identification decreases according to the increase of a sample size as follows.

\begin{defn}\label{def:sic}
    Let $N$ be a sample size for system identification by an adversary.
    A sample identifying complexity $\gamma$ is defined as a function satisfying $\gamma(N)\le\EV[\epsilon]$, where $\epsilon$ is an estimation error of the system identification.
\end{defn}

A sample deciphering time is defined based on a computation time $\tau$ for breaking an encryption scheme offering $\lambda$-bit security can be estimated by $\tau=2^\lambda/\Upsilon$ if an attacker employs a computer of which performance is $\Upsilon$ floating point number operations per second (FLOPS).

\begin{defn}\label{def:sdt}
    A sample deciphering time $\tau$ is a computation time required for breaking $N$ ciphertexts used for system identification by an adversary defined as $\tau(N,1^\lambda)=2^\lambda N/\Upsilon$.
\end{defn}

The security of encrypted control systems is defined using the sample identifying complexity and sample deciphering time as follows.

\begin{defn}\label{def:security}
    Let $\gamma_c$ be an acceptable estimation error, and $\tau_c$ be a life span of plant.
    An encrypted control system is secure if there does not exist $N$ such that $\gamma(N)<\gamma_c$ and $\tau(N,1^\lambda)\le\tau_c$, where $\gamma$ and $\tau$ are defined in \defref{def:sic} and \defref{def:sdt}, respectively.
    Otherwise, the encrypted control system is unsecure.
\end{defn}

\begin{rem}
    The sample deciphering time in \defref{def:sdt} is defined for encrypted control systems using dynamic-key encryption, of which keys are updated every time step~\citep{teranishi2022}.
    In the case of using a traditional homomorphic encryption scheme, the sample deciphering time is computed as $\tau(1,1^\lambda)$ because an adversary can obtain the original message of any ciphertext once the encryption scheme is broken.
\end{rem}

\section{Attack Model}\label{sec:attack_model}

\begin{figure}[t]
    \centering
    \subfigure[Eavesdropper.]{\includegraphics[scale=.9]{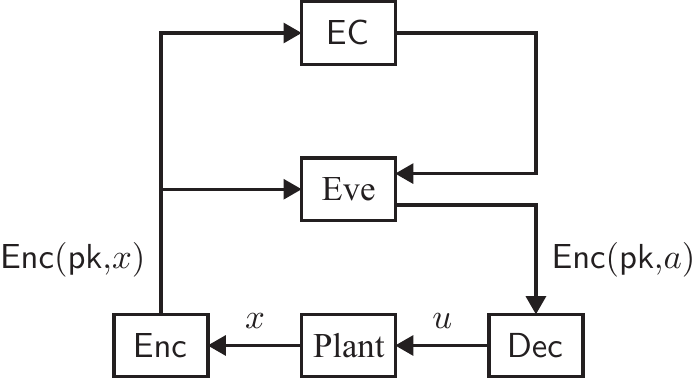}\label{fig:sia_eavesdropper}}
    \subfigure[Malicious server.]{\includegraphics[scale=.9]{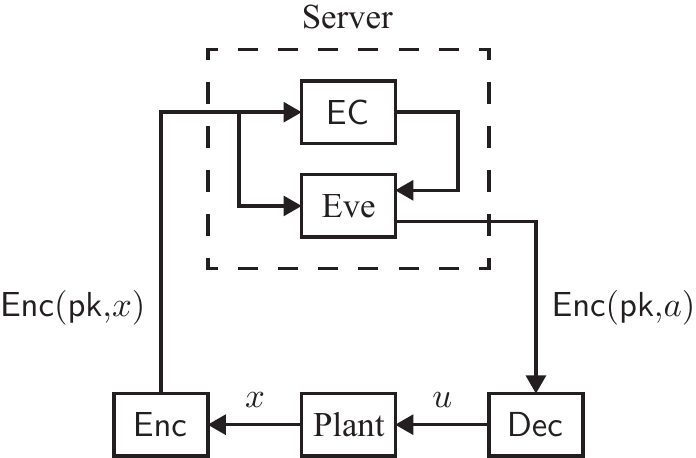}\label{fig:sia_server}}
    \caption{Two types of Adversaries identifying the plant.}
    \label{fig:adversary}
\end{figure}

This section formulates an attack model considered in this study.
\figref{fig:adversary} shows two types of adversaries that aim to identify plant parameters.
Eve in \figref{fig:sia_eavesdropper} is an adversary eavesdropping on network signals and exploiting illegal input signals to a communication channel from the encrypted controller to the decryptor.
This type of adversary represents man-in-the-middle attacks.
\figref{fig:sia_server} depicts another adversary performing system identification.
In the figure, Eve is in a server computing an encrypted control algorithm.
The server records inputs of the encrypted controller algorithm and returns falsified outputs.
Thus, it is called a malicious server that represents a server infected by malware or spoofing as an authorized agent.
It should be noted here that the signal flow of encrypted control systems under the adversaries in \figref{fig:adversary} is the same structure.
Hence, we can deal with the attacks by a unified attack model without assuming the adversary types.

Suppose the plant in \figref{fig:adversary} is give as
\begin{equation}
    x_{t+1} = A x_t + B u_t + w_{t+1},
    \label{eq:plant}
\end{equation}
where $x_0=w_0$, $t\in\Z^+$ is a time, $x\in\R^n$ is a state, $u\in\R^m$ is an input, and $w$ is an i.i.d. random noise following the Gaussian distribution with mean $\bm{0}$ and variance $\sigma_w^2I$.
$A$ and $B$ are system parameters, and $A$ is assumed to be stable.

This study analyzes a sample identifying complexity of \eqref{eq:plant} under an adversary following the protocol: 1) collecting some encrypted samples, 2) exposing the original data by breaking the samples, and 3) identifying system parameters $(A,B)$ by a least squares method with the exposed data.
The attack scenario is formally defined as follows.

\begin{defn}\label{def:adversary}
    The adversary attempts to identify $(A,B)$ of \eqref{eq:plant} by the following procedure.
    \begin{enumerate}
        \item The adversary injects malicious inputs $u_t=a_t$ for $t\in[0,j]$ and collects $N=j+1$ ciphertexts of inputs and states $\{(\ct_{u_k},\ct_{x_k})\}_{k=0}^j$, where $\ct_{u_k}=\Enc(\pk,u_k)$, and $\ct_{x_k}=\Enc(\pk,x_k)$.
        \item The adversary exposes $\{(u_k,x_k)\}_{k=0}^j$ deciphering the ciphertexts.
        \item Then, the adversary estimates $(A,B)$ by a least squares method with the exposed data.
    \end{enumerate}
\end{defn}

\begin{rem}
    In the first step of \defref{def:adversary}, the malicious inputs $a_t$ can be injected properly even though control inputs are encrypted by homomorphic encryption because, in general, an encryption scheme and public key are published.
\end{rem}

For the third step in \defref{def:adversary}, we employ the following least squares identification method.
Define data matrices
\begin{alignat*}{2}
    X_{p} &=
    \begin{bmatrix}
        x_0 & \cdots & x_{j-1}
    \end{bmatrix}, &\quad
    X_f &=
    \begin{bmatrix}
        x_1 & \cdots & x_j
    \end{bmatrix}, \\
    U_p &=
    \begin{bmatrix}
        u_0 & \cdots & u_{j-1}
    \end{bmatrix}, &\quad
    W_f &=
    \begin{bmatrix}
        w_1 & \cdots & w_j
    \end{bmatrix}.
\end{alignat*}
It follows from \eqref{eq:plant} that
\begin{equation}
    X_f = A X_p + B U_p + W_f =
    \begin{bmatrix}
        A & B
    \end{bmatrix}
    \begin{bmatrix}
        X_p \\
        U_p
    \end{bmatrix} + W_f.
    \label{eq:data}
\end{equation}
The least squares estimators $(\hat{A},\hat{B})$ of $(A,B)$ are given as
\begin{equation}
    \begin{bmatrix}
        \hat{A} & \hat{B}
    \end{bmatrix} \!=\!
    \argmin_{[A\ B]} \left\| X_f -
    \begin{bmatrix}
        A & B
    \end{bmatrix}
    \begin{bmatrix}
        X_p \\
        U_p
    \end{bmatrix}
    \right\|_F^2 \!=\! X_f
    \begin{bmatrix}
        X_p \\
        U_p
    \end{bmatrix}^+\!,
    \label{eq:estimate}
\end{equation}
where $([X_p^\top\ U_p^\top]^\top)^+$ is the pseudo inverse matrix of $[X_p^\top\ U_p^\top]^\top$.

\section{Sample Identifying Complexity}
\label{sec:sic}

A sample identifying complexity and sample deciphering time are crucial for defining the security of encrypted control systems in \defref{def:security}.
The sample deciphering time in \defref{def:sdt} can be computed by determining a security parameter and computer performance of the adversary.
In contrast, a computation method for a sample identifying complexity is not obvious because it depends on system dynamics and a system identification method.
This section proposes a sample identifying complexity of \eqref{eq:plant} under the adversary in \defref{def:adversary}.
To this end, we define an estimation error of the least squares identification method as follows.

\begin{defn}\label{def:error}
    An estimation error $\epsilon$ of \eqref{eq:estimate} is defined as
    \[
        \epsilon = \cfrac{1}{c} \left\|
        \begin{bmatrix}
            A & B
        \end{bmatrix} -
        \begin{bmatrix}
            \hat{A} & \hat{B}
        \end{bmatrix}
        \right\|_F^2,
    \]
    where $c=n(n+m)$ is the number of entries of $A$ and $B$.
\end{defn}

By \defref{def:error}, $\epsilon$ is a mean square error of the estimates $\hat{A}$ and $\hat{B}$.
It should be noted here that one of the best strategies for the adversary in \defref{def:adversary} to design the malicious inputs $a_0,\dots,a_j$ minimizing the error $\epsilon$ is that the inputs are independently and identically sampled from the Gaussian distribution with mean zero.
Under this setting, the following theorem reveals a lower bound for the expectation of estimation error.

\begin{thm}\label{thm:sic}
    Let $N=j+1$.
    Suppose malicious inputs $a_0,\dots,a_j$ are i.i.d. noises following the Gaussian distribution with mean $\bm{0}$ and variance $\sigma_u^2I$.
    The expectation of estimation error in \defref{def:error} is bounded from below by
    \begin{equation}
        \EV\left[ \epsilon \right] \!\ge\! \gamma(N) \!\coloneqq\! \cfrac{(m + n) \sigma_w^2}{j \left( m + \tr(\Psi_1) \right) \sigma_u^2 + \left( j n + \tr(\Psi_2) \right) \sigma_w^2},
        \label{eq:sic}
    \end{equation}
    where $\Psi_1$ and $\Psi_2$ are controllability Gramians respectively obtained by solving the following discrete Lyapunov equations,
    \begin{align*}
        & A \Psi_1 A^\top - \Psi_1 + B B^\top = O, \\
        & A \Psi_2 A^\top - \Psi_2 + I = O.
    \end{align*}
\end{thm}

\begin{pf}
    Let $D = [X_p^\top\ U_p^\top]^\top$.
    It follows from \eqref{eq:data} and \eqref{eq:estimate} that
    \begin{align*}
            \EV[\epsilon]
        &=  \cfrac{1}{c} \EV\left[ \left\| 
            \begin{bmatrix}
                A & B
            \end{bmatrix}
            - X_f D^+ \right\|_F^2 \right], \\
        &=  \cfrac{1}{c} \EV\left[ \left\|
            \begin{bmatrix}
                A & B
            \end{bmatrix}
            - \left(
            \begin{bmatrix}
                A & B
            \end{bmatrix}
            D + W_f \right) D^+ \right\|_F^2 \right], \\
        &=  \cfrac{1}{c} \EV\left[ \left\| W_f D^+ \right\|_F^2 \right], \\
        &=  \cfrac{1}{c} \EV\left[ \left\| \Vec( W_f D^+ ) \right\|_2^2 \right], \\
        &=  \cfrac{1}{c} \EV\left[ \tr\left( \Vec( W_f D^+ ) \Vec( W_f D^+ )^\top \right) \right], \\
        &=  \cfrac{1}{c} \EV\left[ \tr\left( \left( D^+ \!\otimes\! I \right)^\top \Vec(W_f) \Vec(W_f)^\top \left( D^+ \!\otimes\! I \right) \right) \right], \\
        &=  \cfrac{1}{c} \tr\left( \EV\left[ D^+ (D^+)^\top \otimes I \right] \EV\left[ \Vec(W_f) \Vec(W_f)^\top \right] \right),
    \end{align*}
    where $\otimes$ is the Kronecker product.
    Additionally,
    \[
          \EV\left[ \Vec(W_f) \Vec(W_f)^\top \right]
        = \EV\left[
            \begin{bmatrix}
                w_1 w_1^\top & \cdots & w_1 w_j^\top \\
                \vdots       & \ddots & \vdots       \\
                w_j w_1^\top & \cdots & w_j w_j^\top
            \end{bmatrix}
          \right]
        = \sigma_w^2 I.
    \]
    Hence, we obtain
    \begin{align*}
           \EV[\epsilon]
        &= \cfrac{\sigma_w^2}{c} \tr\left( \EV\left[ D^+ (D^+)^\top \otimes I \right] \right), \\
        &= \cfrac{\sigma_w^2}{c} \tr\left( \EV\left[ D^+ (D^+)^\top \right] \right) \tr(I), \\
        &= \cfrac{n \sigma_w^2}{c} \EV\left[ \tr\left( D^\top (D D^\top)^{-1} (D D^\top)^{-1} D \right) \right], \\
        &= \cfrac{n \sigma_w^2}{c} \EV\left[ \tr\left( (D D^\top)^{-1} \right) \right].
    \end{align*}
    Using Jensen's inequality, the expectation of trace of inverse matrix is bounded from below by
    \begin{align*}
             \EV\left[ \tr\left( (D D^\top)^{-1} \right) \right] 
        &\ge (m + n)^2 \EV\left[ \tr\left( D D^\top \right)^{-1} \right], \\
        &\ge (m + n)^2 \EV\left[ \tr\left( D D^\top \right) \right]^{-1}, \\
        &=   (m + n)^2 \tr\left( \EV\left[ D D^\top \right] \right)^{-1}.
    \end{align*}
    The denominator is computed as
    \begin{align*}
            \tr\left( \EV\left[ D D^\top \right] \right)
        &=  \tr\left( \EV\left[
            \begin{bmatrix}
                X_p X_p^\top & X_p U_p^\top \\
                U_p X_p^\top & U_p U_p^\top
            \end{bmatrix}
            \right] \right) \\
        &=  \tr\left( \EV\left[ X_p X_p^\top \right] \right) + \tr\left( \EV\left[ U_p U_p^\top \right] \right) \\
        &=  \tr\left( \EV\left[
            \begin{bmatrix}
                x_0 & \cdots & x_{j-1}
            \end{bmatrix}
            \begin{bmatrix}
                x_0^\top     \\
                \vdots       \\
                x_{j-1}^\top \\
            \end{bmatrix}
            \right] \right) \\
        &\quad+  \tr\left( \EV\left[
            \begin{bmatrix}
                u_0 & \cdots & u_{j-1}
            \end{bmatrix}
            \begin{bmatrix}
                u_0^\top     \\
                \vdots       \\
                u_{j-1}^\top \\
            \end{bmatrix}
            \right] \right), \\
        &=  \tr\left( \EV\left[ \sum_{t=0}^{j-1} x_t x_t^\top \right] \right) \!+\! \tr\left( \EV\left[ \sum_{t=0}^{j-1} u_t u_t^\top \right] \right).
    \end{align*}
    It follows from \eqref{eq:plant} that
    \[
        x_t = A^t x_0 + \sum_{k=0}^{t-1} A^{t-1-k} B u_k + w_t.
    \]
    Thus, the traces of expectations are given as
    \begin{align*}
        &       \tr\left( \EV\left[ \sum_{t=0}^{j-1} x_t x_t^\top \right] \right) \\
        &=      \tr\left( \EV\left[ \sum_{t=0}^{j-1} A^t x_0 x_0^\top (A^t)^\top \right] \right) \\
        &\quad+ \tr\left( \EV\left[ \sum_{t=0}^{j-1} \sum_{k=0}^{t-1} A^{t-1-k} B u_k u_k^\top B^\top (A^{t-1-k})^\top \right] \right) \\
        &\quad+ \tr\left( \EV\left[ \sum_{t=0}^{j-1} w_t w_t^\top \right] \right), \\
        &=      j n \sigma_w^2 \!+\! \sigma_w^2 \tr\!\left( \sum_{t=0}^{j-1} \!A^t (A^t)^{\!\top} \!\!\right) \!\!+\! \tr\!\left( \sum_{t=0}^{j-1} \!\sum_{k=0}^{t-1} \!A^k B B^\top \!(A^k)^{\!\top} \!\!\right)\!, \\
        &       \tr\left( \EV\left[ \sum_{t=0}^{j-1} u_t u_t^\top \right] \right) = \sum_{t=0}^{j-1} \sigma_u^2 \tr(I) = j m \sigma_u^2.
    \end{align*}
    Furthermore, the matrices are bounded by
    \begin{align*}
        & \sum_{k=0}^{t-1} A^k B B^\top (A^k)^\top \le \sum_{k=0}^\infty A^k B B^\top (A^k)^\top = \Psi_1, \\
        & \sum_{t=0}^{j-1} A^t (A^t)^\top \le \sum_{t=0}^\infty A^t (A^t)^\top = \Psi_2.
    \end{align*}
    Therefore, we obtain the bound for the expectation of estimation error as
    \begin{align*}
                \EV[\epsilon]
        &=      \cfrac{n \sigma_w^2}{n (n + m)} \cdot (m + n)^2 \cdot \cfrac{1}{\tr\left( \EV\left[ D D^\top \right] \right)} \\
        &\ge    \cfrac{(m + n) \sigma_w^2}{j \left( m + \tr(\Psi_1) \right) \sigma_u^2 + \left( j n + \tr(\Psi_2) \right) \sigma_w^2}.
    \end{align*}
    This completes the proof. \qed
\end{pf}

From \thmref{thm:sic}, the lower bound $\gamma$ in \eqref{eq:sic} is a sample identifying complexity of \eqref{eq:plant} under the adversary in \defref{def:adversary}.
Now, if $j$ is sufficiently large, $\gamma$ can be approximated by a simple equation.

\begin{cor}
    Let $N=j+1$ and $R_\sigma=\sigma_u^2/\sigma_w^2$.
    Suppose $j$ is sufficiently large.
    Then, the sample identifying complexity is given as
    \begin{equation}
        \gamma(N) = \cfrac{m+n}{j \left[ (m + \tr(\Psi_1)) R_\sigma + n \right]}.
        \label{eq:sic2}
    \end{equation}
\end{cor}

\begin{pf}
    If $j$ is sufficiently large, the denominator of \eqref{eq:sic} is approximated as $j\left(m+\tr(\Psi_1)\right)\sigma_u^2+jn\sigma_w^2$.
    Then, \eqref{eq:sic2} holds by dividing both the numerator and denominator of \eqref{eq:sic} by $\sigma_w^2$. \qed
\end{pf}

The equation \eqref{eq:sic2} shows that the sample identifying complexity is characterized by the eigenvalues of controllability Gramian $\Psi_1$ and variance ratio $R_\sigma$.
If $R_\sigma$ is small, the term $\tr(\Psi_1)R_\sigma$ becomes small.
Then, the sample identifying complexity $\gamma$ is almost independent of the eigenvalues of controllability Gramian.
Meanwhile, the smaller eigenvalues of the controllability Gramian are, the larger sample identifying complexity is.
In other words, the expectation of estimation error in \defref{def:error} is larger if the stability degree of plant \eqref{eq:plant} becomes larger.
This implies that, when $R_\sigma$ is large, the information leakage on the dynamics of \eqref{eq:plant} can be reduced by making \eqref{eq:plant} more stable because a plant state is not sufficiently driven by an input for system identification.
Note that a similar observation has been discussed in the previous study in which the adversary aims to identify a system matrix of a closed-loop system using the Bayesian estimation~\citep{teranishi2022}.

Now we consider extending the sample identifying complexity to a closed-loop case.
With the plant \eqref{eq:plant} and an encrypted controller $\EC(f_\mathrm{sf},\ct_F,\ct_r,\ct_x)$ of state feedback controller $f_\mathrm{sf}: (F, x_t) \mapsto u_t = F x_t + r_t$, the closed-loop system is given as $x_{t+1} = A_F x_t + B r_t + w_{t+1}$, where $A_F=A+BF$, $F$ is a feedback gain, and $r\in\R^m$ is a reference input to be falsified by an adversary.
We assume that an adversary focuses on identifying the parameters $A_F$ and $B$ by a least squares method in \secref{sec:attack_model} with the data $\{(r_k,x_k)\}_{k=0}^j$, which is obtained by deciphering $\{(\ct_{r_k},\ct_{x_k})\}_{k=0}^j$.
Then, a controllability Gramian $\Psi_1$ is given by solving the discrete Lyapunov equation $A_F\Psi_1A_F^\top-\Psi_1+BB^\top=O$, and explicitly formulated as $\Psi_1=\sum_{k=0}^\infty A_F^kBB^\top(A_F^k)^\top$.
Here the controllability Gramian is a function of the feedback gain $F$.
Thus, we would be able to improve $\gamma$ of the closed-loop system by designing the feedback gain so that eigenvalues of the controllability Gramian are minimized.
Note that this defense policy would be effective only when the variance ratio is sufficiently large since the effect of controllability Gramian for the sample identifying complexity can be insignificant if the ratio is small.

\section{Numerical Simulation}
\label{sec:simulation}

\begin{figure}[t]
    \centering
    \includegraphics[scale=.9]{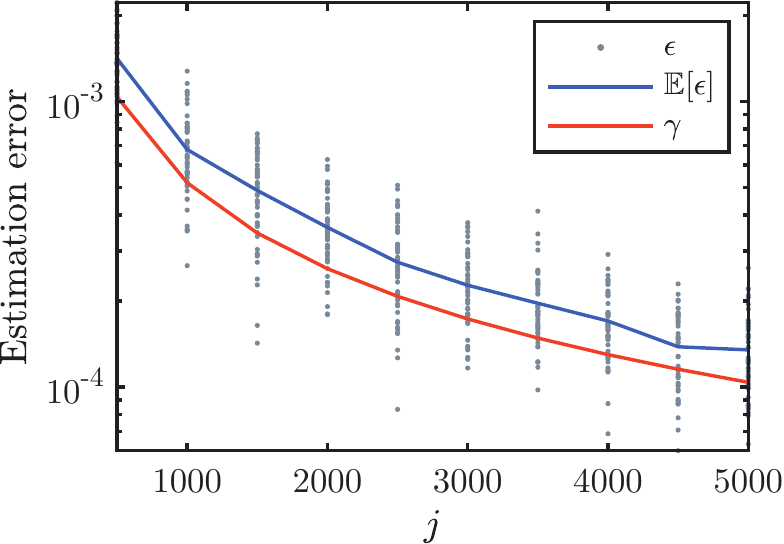}
    \caption{Comparison between the expectation of estimation error and sample identifying complexity.}
    \label{fig:err}
\end{figure}

Consider the plant \eqref{eq:plant} with the parameters
\[
    A = 
    \begin{bmatrix}
         0.23 &  0.45 & -0.04 & -0.04 \\
         0.45 & -0.46 & -0.12 &  0.15 \\
        -0.04 & -0.12 &  0.43 & -0.02 \\
        -0.04 &  0.15 & -0.02 &  0.20
    \end{bmatrix}, \ 
    B = 
    \begin{bmatrix}
         0.27 & -1.32 \\
        -0.29 & -0.31 \\
        -0.64 &  0.75 \\
        -0.13 & -0.97
    \end{bmatrix}.
\]
The variances are set to $\sigma_w^2=\sigma_u^2=1$.
\figref{fig:err} shows a comparison between the expectation of estimation error and sample identifying complexity of the plant.
The system identification is performed $50$ times for each sample size with different data sets based on the dynamics of \eqref{eq:plant}.
The gray dots in the figure are the estimation errors in \defref{def:error}.
The blue and red lines are the expectation of estimation error and the sample identifying complexity \eqref{eq:sic}, respectively.
The result shows that the proposed complexity captures the behavior of expected error with a sufficient level.

\begin{figure}[t]
    \centering
    \subfigure[$\sigma_w^2\!=\!0.1,\sigma_u^2\!=\!0.1$.]{\includegraphics[scale=.36]{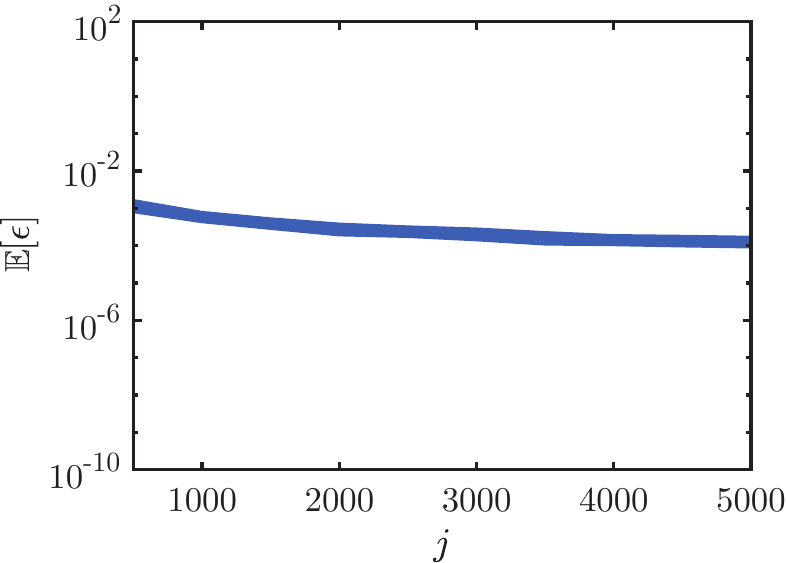}\label{fig:w01_u01_ex}}%
    \subfigure[$\sigma_w^2\!=\!0.1,\sigma_u^2\!=\!1$.]{\includegraphics[scale=.36]{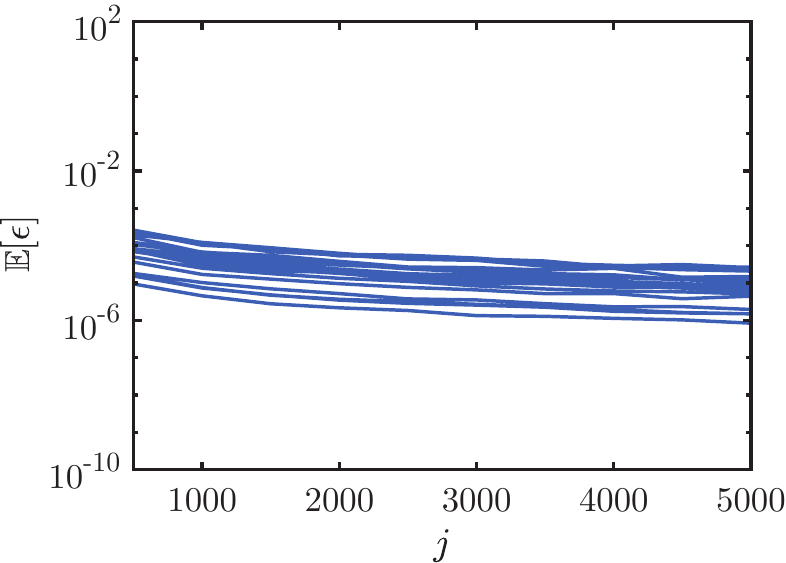}\label{fig:w01_u1_ex}}%
    \subfigure[$\sigma_w^2\!=\!0.1,\sigma_u^2\!=\!10$.]{\includegraphics[scale=.36]{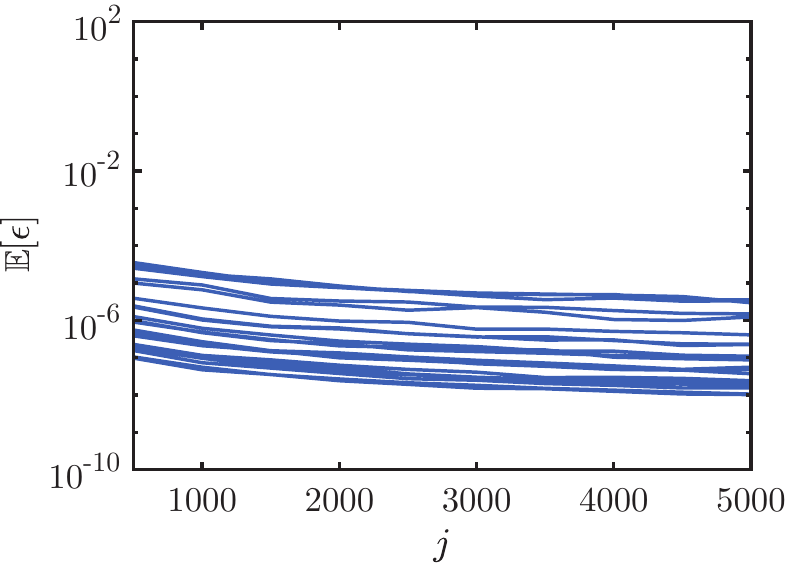}\label{fig:w01_u10_ex}}
    \subfigure[$\sigma_w^2\!=\!1,\sigma_u^2\!=\!0.1$.]{\includegraphics[scale=.36]{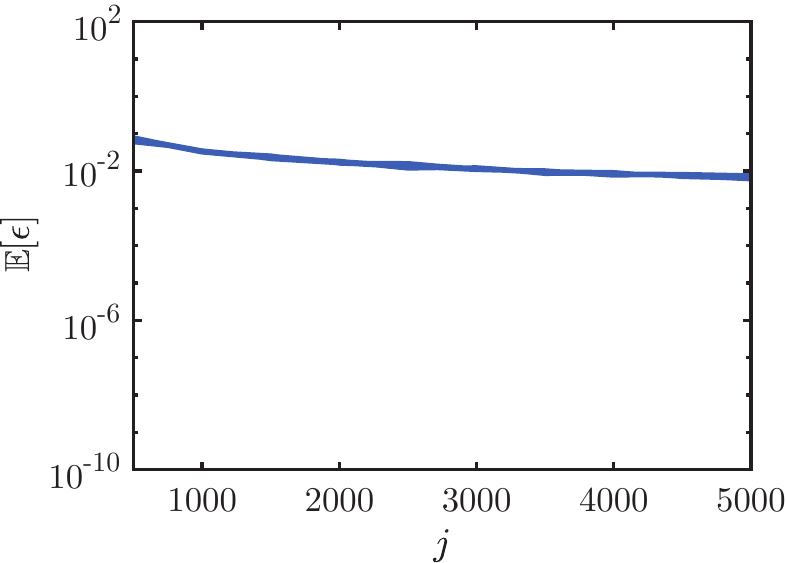}\label{fig:w1_u01_ex}}%
    \subfigure[$\sigma_w^2\!=\!1,\sigma_u^2\!=\!1$.]{\includegraphics[scale=.36]{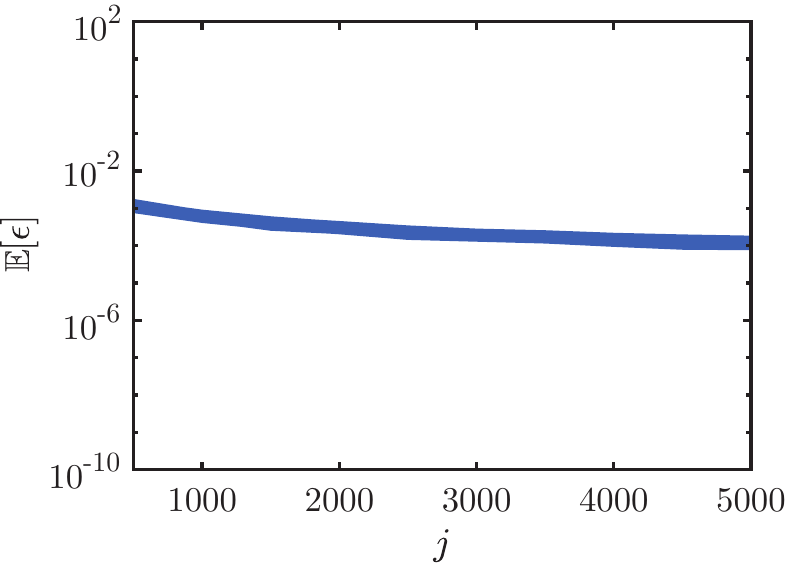}\label{fig:w1_u1_ex}}%
    \subfigure[$\sigma_w^2\!=\!1,\sigma_u^2\!=\!10$.]{\includegraphics[scale=.36]{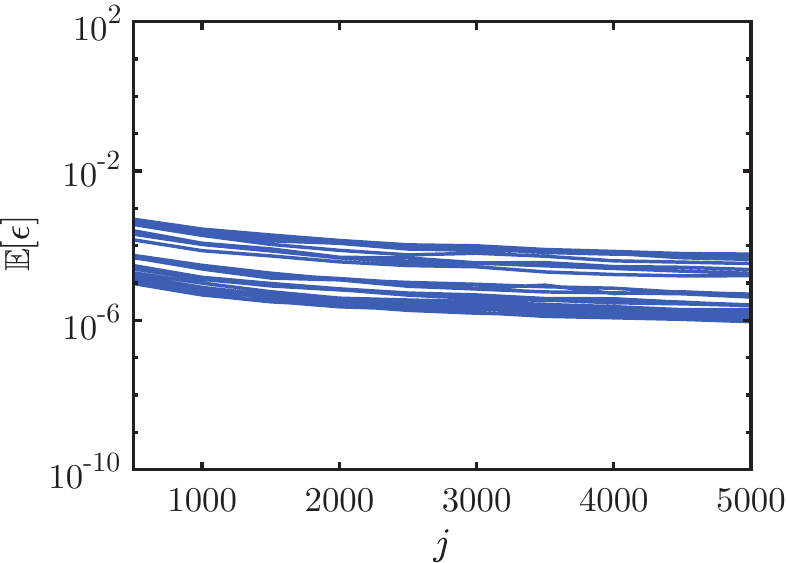}\label{fig:w1_u10_ex}}
    \subfigure[$\sigma_w^2\!=\!10,\sigma_u^2\!=\!0.1$.]{\includegraphics[scale=.36]{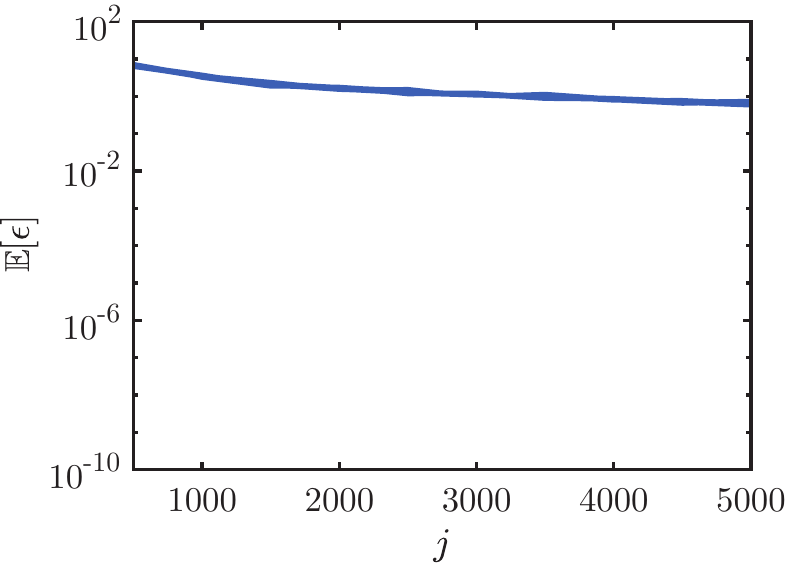}\label{fig:w10_u01_ex}}%
    \subfigure[$\sigma_w^2\!=\!10,\sigma_u^2\!=\!1$.]{\includegraphics[scale=.36]{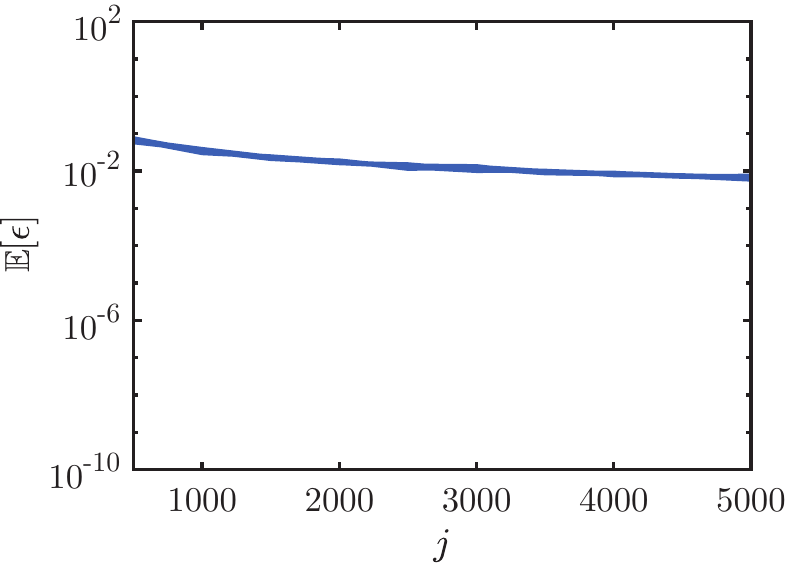}\label{fig:w10_u1_ex}}%
    \subfigure[$\sigma_w^2\!=\!10,\sigma_u^2\!=\!10$.]{\includegraphics[scale=.36]{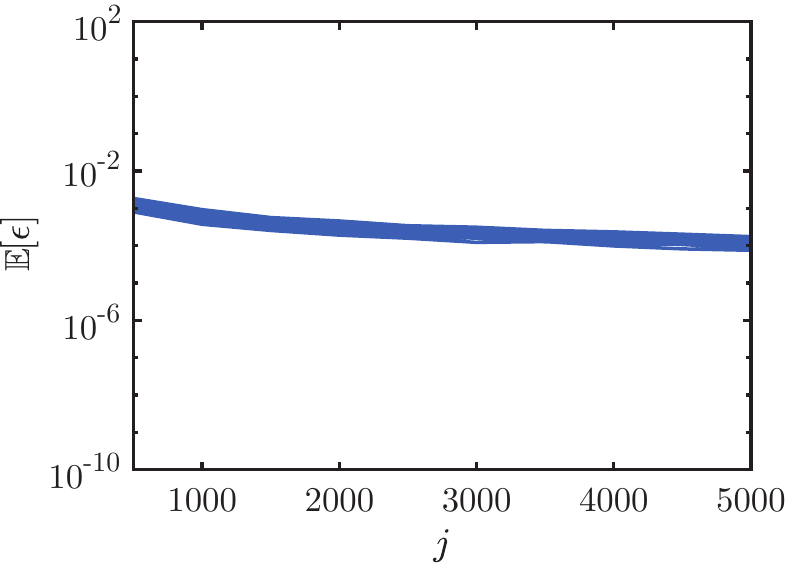}\label{fig:w10_u10_ex}}
    \caption{The expectation of estimation error.}
    \label{fig:ex}
\end{figure}

\begin{figure}[t]
    \centering
    \subfigure[$\sigma_w^2\!=\!0.1,\sigma_u^2\!=\!0.1$.]{\includegraphics[scale=.36]{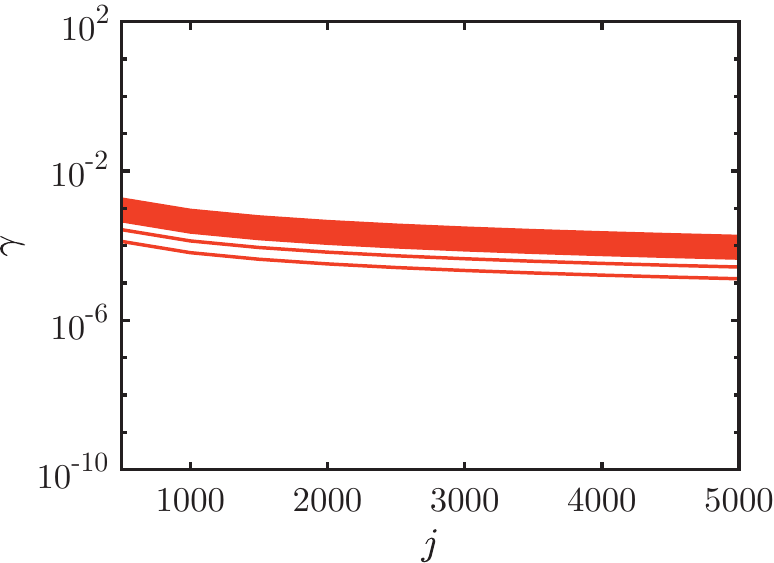}\label{fig:w01_u01_sic}}%
    \subfigure[$\sigma_w^2\!=\!0.1,\sigma_u^2\!=\!1$.]{\includegraphics[scale=.36]{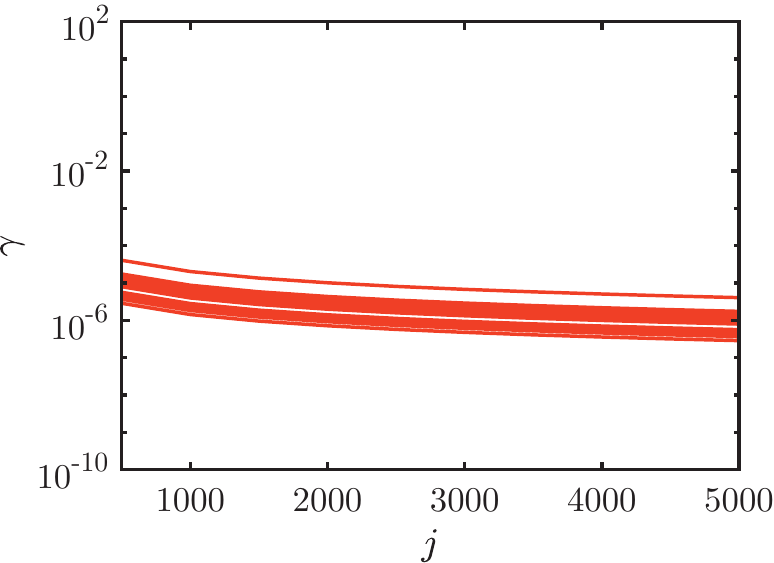}\label{fig:w01_u1_sic}}%
    \subfigure[$\sigma_w^2\!=\!0.1,\sigma_u^2\!=\!10$.]{\includegraphics[scale=.36]{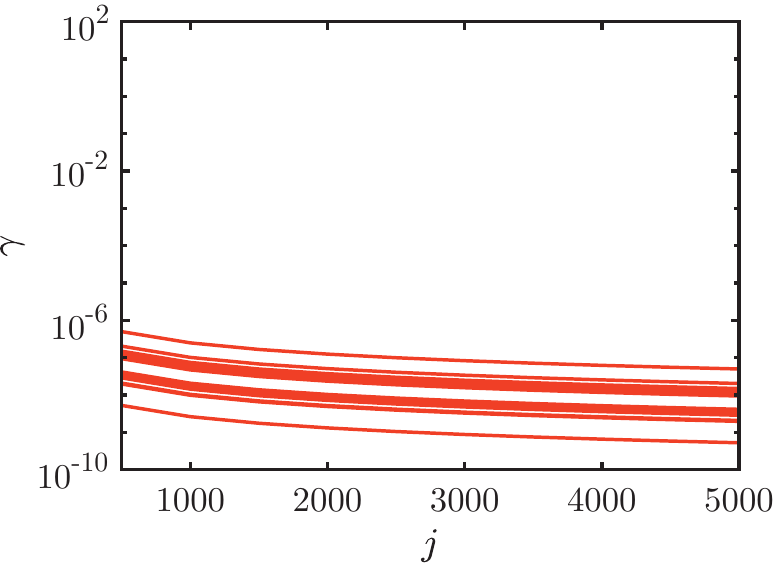}\label{fig:w01_u10_sic}}
    \subfigure[$\sigma_w^2\!=\!1,\sigma_u^2\!=\!0.1$.]{\includegraphics[scale=.36]{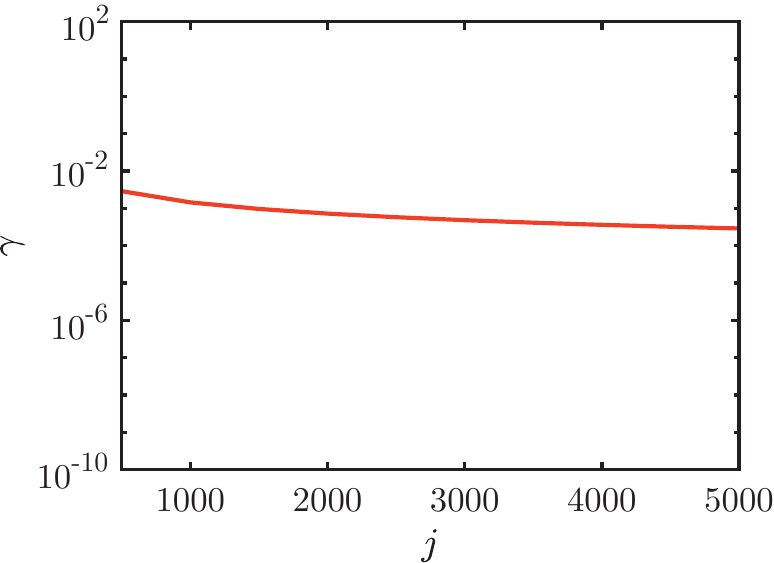}\label{fig:w1_u01_sic}}%
    \subfigure[$\sigma_w^2\!=\!1,\sigma_u^2\!=\!1$.]{\includegraphics[scale=.36]{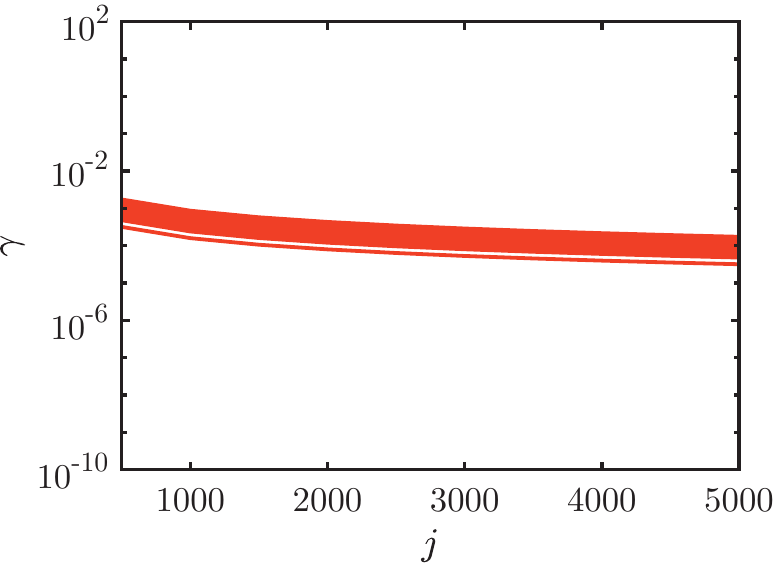}\label{fig:w1_u1_sic}}%
    \subfigure[$\sigma_w^2\!=\!1,\sigma_u^2\!=\!10$.]{\includegraphics[scale=.36]{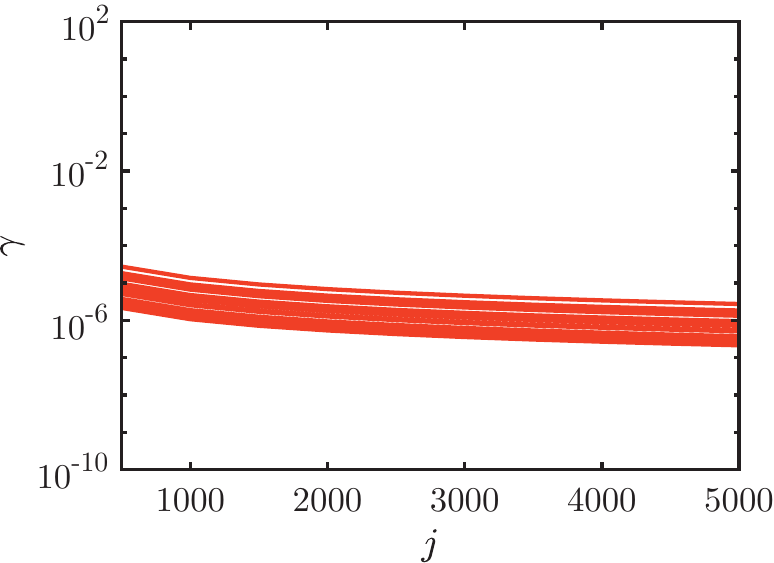}\label{fig:w1_u10_sic}}
    \subfigure[$\sigma_w^2\!=\!10,\sigma_u^2\!=\!0.1$.]{\includegraphics[scale=.36]{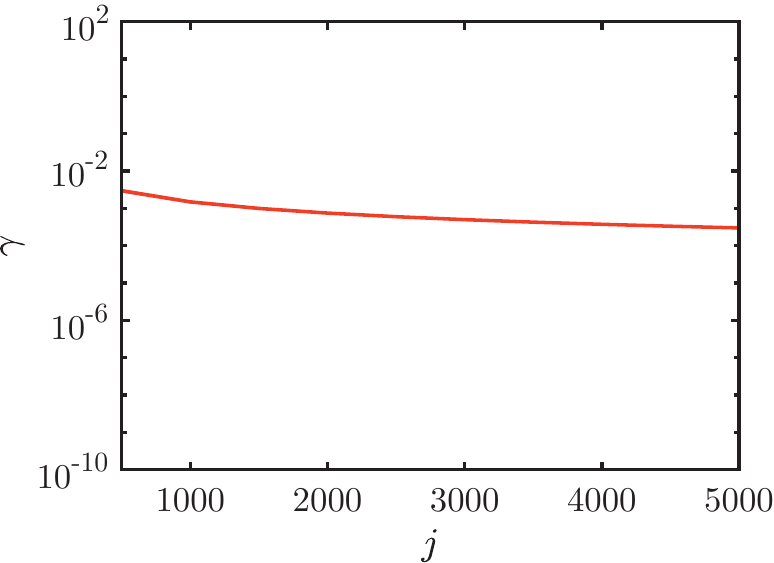}\label{fig:w10_u01_sic}}%
    \subfigure[$\sigma_w^2\!=\!10,\sigma_u^2\!=\!1$.]{\includegraphics[scale=.36]{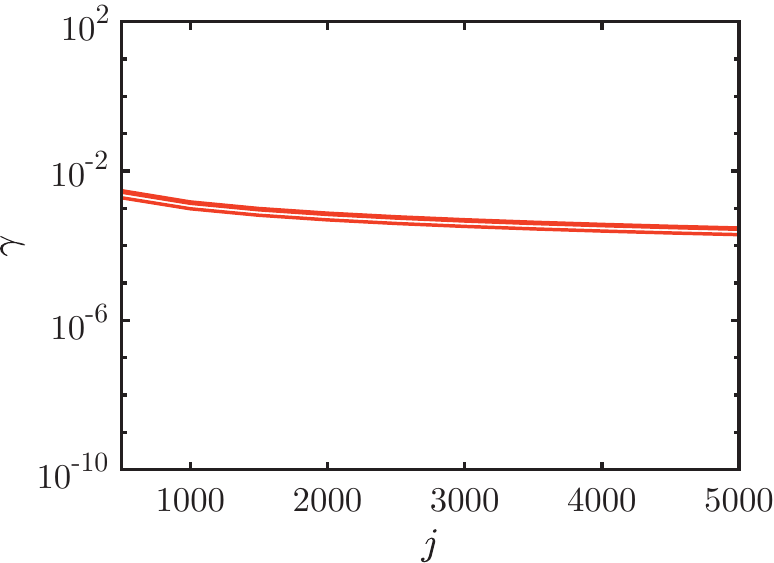}\label{fig:w10_u1_sic}}%
    \subfigure[$\sigma_w^2\!=\!10,\sigma_u^2\!=\!10$.]{\includegraphics[scale=.36]{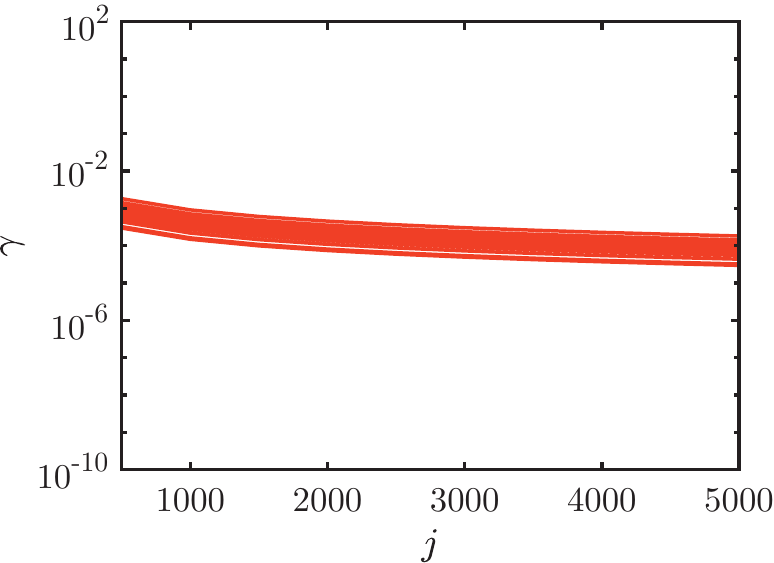}\label{fig:w10_u10_sic}}
    \caption{Sample identifying complexity.}
    \label{fig:sic}
\end{figure}

Next, we confirm changes in the expectation of estimation error and sample identifying complexity when the variances are varied.
\figref{fig:ex} shows the expectations of estimation errors with the nine combinations of $\sigma_w^2=0.1,1,10$ and $\sigma_u^2=0.1,1,10$.
The expected errors are computed for each setting using data sets obtained by $50$ trials for $20$ plants, which are randomly generated.
From the figures, the expected errors tend to decrease as the variance ratio $R_\sigma=\sigma_u^2/\sigma_w^2$ increases.
The sample identifying complexities shown in \figref{fig:sic}, which are computed with the same settings of \figref{fig:ex}, depict similar behaviors as the expected errors.
These results confirm that the proposed complexity properly represents a change in an expected error according to a variance ratio.
Moreover, it can be seen from \figref{fig:ex} and \figref{fig:sic} that the variation of estimated error and sample identifying complexity is larger as the variance ratio increases.
These results imply that the effect of controllability Gramian on the error and complexity become larger as the variance ratio increases.

\begin{figure}[t]
    \centering
    \includegraphics[scale=.9]{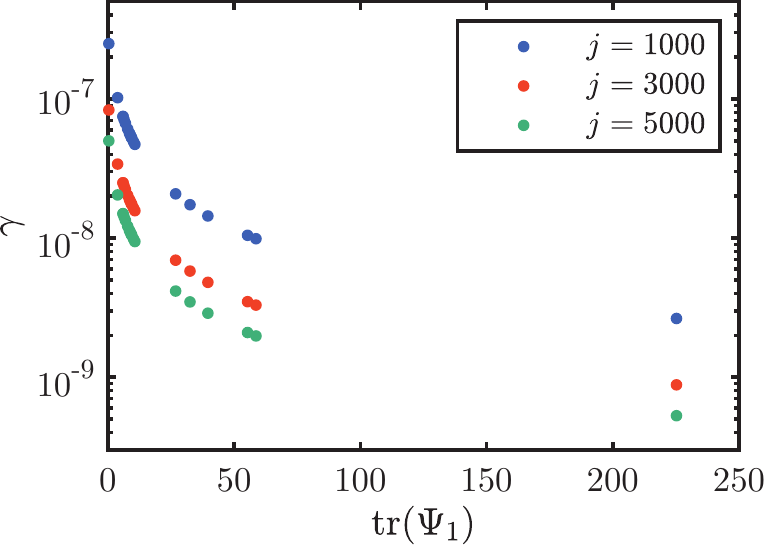}
    \caption{Change in the sample identifying complexity associated with the eigenvalues of controllability Gramian.}
    \label{fig:tr_sic}
\end{figure}

Finally, we examine a change in the sample identifying complexity associated with the eigenvalues of controllability Gramian for the same setting in \figref{fig:sic}\subref{fig:w01_u10_sic}.
Note that the variance ratio in this case is $R_\sigma=100$ ($\sigma_w^2=0.1$, $\sigma_u^2=10$).
\figref{fig:tr_sic} shows the sample identifying complexities with $j=1000,3000,5000$ for the $20$ variations of the trace of controllability Gramian, which are computed using the randomly generated plants.
For all sample sizes, it can be seen from the figure that the smaller $\tr(\Psi_1)$ is, the larger $\gamma$ is.
These results mean that the sample identifying complexity can be improved by modifying eigenvalues of the controllability Gramian.

\section{Conclusion}
\label{sec:conclusion}

This study proposed a sample identifying complexity under an adversary who tries to identify parameters of a plant in an encrypted control system using a least squares method.
The proposed sample identifying complexity is computed by system dimensions, controllability Gramians, and noise and input variances.
The simulation results demonstrated that the proposed complexity captures the expectation of estimation error with a sufficient level.
Our future work includes deriving a sample identifying complexity of multi-agent and nonlinear systems.

\bibliography{encrypted_control_and_optimization,others}

\end{document}